# Pressure-induced decomposition of solid hydrogen sulfide

Defang Duan, Xiaoli Huang, Fubo Tian, Da Li, Hongyu, Yu, Yunxian Liu, Yanbin Ma, Bingbing Liu, Tian Cui [a]

*State Key Laboratory of Superhard Materials, College of Physics,*
*Jilin University, Changchun 130012, People's Republic of China*

[a] Electronic address: cuitian@jlu.edu.cn

Solid hydrogen sulfide is well known as a typical molecular crystal but its stability under pressure is still under debate. Particularly, Eremets et al. found the high pressure superconductivity with $T_c \approx 190$ K in a $H_2S$ sample [arXiv: 1412.0460 (2014)] which is associates with the elemental decomposition into $H_3S$ [Sci. Rep. 4, 6968 (2014)]. Therefore, on what pressure $H_2S$ can decompose and which kind of the products of decomposition urgent need to be solved. In this paper, we have performed an extensive structural study on different stoichiometries $H_nS$ with n> 1 under high pressure using *ab initio* calculations. Our results show that $H_2S$ is stable below 50 GPa and decomposes into $H_3S$ and sulfur at high pressure, while $H_3S$ is stable at least up to 300 GPa. The other hydrogen-rich $H_4S$, $H_5S$, and $H_6S$ are unstable in the pressure range from 20 to 300 GPa.

# I. Introduction

Solid hydrogen sulfide is known as a typical molecular crystal which has been studied extensively at high pressure[1-15]. Up to now, the exact structures and molecular dissociation of $H_2S$ at high-pressure is still under debate. Experimentally, at room temperature, hydrogen sulfide transforms to phase IV with monoclinic *Pc*[8] or tetragonal *I4$_1$/acd*[9]. Upon compression, further transformation is complicated by the partial dissociation of $H_2S$ and the appearance of elemental sulfur. But there have been two opinions of the dissociation pressure of $H_2S$ derived from optical experiments. One is 46 GPa[6] at room temperature by infrared-absorption spectra measurement, another is 27 GPa at room temperature and higher pressures around 43 GPa at 150 K by powder x-ray diffraction experiments[13]. Another Raman and IR spectra of $D_2S$ revealed that it dissociates to sulfur at pressures above 27 GPa at room temperature. When the pressure is higher, metallization occur near 96 GPa[6], which can be attribute to elemental sulfur that is known to become metallic above 95 GPa[16].

Theoretically, Rousseau et al. studied the structure and stability of $H_2S$ at room temperature through *ab initio* MD simulations. They suggested that phase IV has partial rotational disorder with *P4$_2$/ncm*[10] and the $H_2S$ molecules lose their identity in phase V with forming $H_3S^+$ and $HS^-$ ionic species and fluctuating S–S chains[11]. Another *ab initio* MD simulation predicted an *Ibca* structure at 15 GPa and 100 K with the initial configuration of the *P4$_2$/ncm* structure. When temperature increases to 350 K, it transforms to a proton disorder structure which is important to the decomposition of $H_2S$. Recent theoretical work have revised the phase diagram of $H_2S$ up to 200 GPa, and suggested that it is stable with respect to decomposition into hydrogen and sulfur. In addition, a superconducting critical temperature ($T_c$) of ~80 K at 160 GPa was predicted.

It is reported experimentally that mixing hydrogen sulfide ($H_2S$) and hydrogen ($H_2$) can form a stoichiometric compound $(H_2S)_2H_2$ near 3.5 GPa, which is a rotationally disordered structure with *I4/mcm* symmetry.[17] Upon compression, an ordering process occurs with structural transformation above 17 GPa. We have explored the high-pressure ordered crystal structures of $(H_2S)_2H_2$ by means of *ab initio* calculations.[18] Four high-pressure phases have been found and the $T_c$ of cubic phase was predicted to reach 191~204 K at 200 GPa for the first time. Recently, Eremets et al. found superconductivity in a $H_2S$ sample with $T_c \approx 190$ K above 150 GPa and 220 K which is likely associates with the dissociation of $H_2S$ into $H_nS$ (n>2) formation[19].

In the present work, we aim to elucidate the high pressure stability of different stoichiometric $H_nS$ (n>1) hydrides using *ab initio* calculations and investigate whether $H_2S$ can decompose and what the products of decomposition are. The results show that $H_2S$ molecule is only stable below 50 GPa and it will decomposes into $H_3S$ and sulfur element at higher pressures.

## II. Methods

Searches for the stable compounds and structures for different stoichiometric $H_nS$ (n>1) hydrides are performed by merging the evolutionary algorithm and *ab* *initio* total-energy calculations, as implemented in the USPEX code (Universal Structure Predictor: Evolutionary Xtallography)[20-22]. The details of the search algorithms and its several applications have been described elsewhere[23,24]. In the first generation, structures (population size: 10-60 structures, increasing with system size) are produced randomly. All produced structures are relaxed and the appropriate thermodynamic potential is used as fitness function. The subsequent generation is created from 60% of the lowest-enthalpy structures of the preceding generation. New structures are created by heredity (60%), permutation (20%), and lattice mutation (20%) operations. The best structure of each generation is also carried over to the next generation. The calculation stops when the best structure does not change for more than 20 generations.

The underlying structure relaxations are performed using density functional theory as implemented in the Vienna *ab initio* simulation package VASP code[25]. The generalized gradient approximation (GGA) of Perdew-Burke-Ernzerhof (PBE)[26] is adopted to describe the exchange-correlation potential. The all-electron projector augmented wave (PAW) method[27] is adopted with the core radii are 0.8 a.u. for H ($1s^2$) and 1.5 a.u. for S ($3s^23p^4$). For the initial search over structures, Brillouin zone (BZ) sampling using a grid of spacing $2\pi \times 0.05$ Å$^{-1}$ and a plane-wave basis set cutoff of 500 eV are found to be sufficient. But we recalculate the enthalpy curves with higher accuracy using the energy cutoff 800 eV and k-mesh of $2\pi \times 0.03$ Å$^{-1}$ within the Monkhorst-Pack (MP) scheme are chosen to ensure that the total energy are well converged to better than 1 meV/atom.

## III. Results and discussion

The structure predictions are performed considering simulation sizes ranging from 1 to 4, 6, and 8 formula units per cell (f. u./cell) for $H_2S$ and $H_3S$ at pressures between 20 and 300 GPa. For the case of $H_4S$, $H_5S$ and $H_6S$, the structure predictions are performed within 2 and 4 f. u. at pressures of 50, 100, 200 and 300 GPa. The stability of $H_nS$ (n>1) can be quantified by constructing the thermodynamic convex hull at the given pressure, which is defined as the formation enthalpy per atom of the most stable phases at each stoichiometry:

$h_f$ ($H_nS$) = [$h$($H_nS$)−$h$(S)−n$h$($H_2$)/2]/(n+1).

where $h_f$ is the enthalpy of formation per atom and $h$ is the calculated enthalpy per chemical unit for each compound. The enthalpies $h$ for $H_nS$ are obtained for the most stable structures as searched by the USPEX method at the desired pressures. The known structures of $P6_3m$, $C2/c$, and $Cmca$ [28] for $H_2$, $I4_1/acd$ [29] and $\beta$-Po [30] for S in their corresponding stable pressure are adopted. Any structure with the enthalpy lies on the convex hull is considered to be thermodynamically stable and synthesizable experimentally[31]. The convex hulls at selected pressure of $H_nS$ are depicted in Fig. 1.

If a tie-line is drawn connecting $h_f$ (A) and $h_f$ (B), while $h_f$ (C) falls beneath it, then compounds A and B will react to form compound C, provided the kinetic barrier is not too high. If another compound D with $h_f$ (D) falls above the tie-line connecting $h_f$ (A) and $h_f$ (B), it is expected to decompose into these two compounds A and B. The solid line in Fig. 1 traces out the estimated convex hull of stoichiometries $H_nS$ (n>1) at 20, 40，50, 100, 200 and 300 GPa. At 20 GPa (Fig. 1a), the enthalpy of formation for $H_2S$ and $H_3S$ stoichiometries fall on the convex hull, indicating that both compounds are thermodynamically stable at this pressure. In addition, $H_2S$ has the most negative enthalpy of formation, which is consistent with the knowledge that $H_2S$ exists at low pressure range. For $H_3S$, it can be synthesized using S and $H_2$ as precursors or using $H_2S$ and $H_2$ as precursors, which has been confirmed by T.A. Strobel et al..[17] They reported that mixtures of $H_2S$ and $H_2$ were loaded into diamond anvil cells and compression up to 3.5 GPa, a compound $(H_2S)_2H_2$ ($H_3S$ with hydrogen-sulfur stoichiometric ratio 3:1) was formed. With increasing pressure, $H_3S$ become the most stable stoichiometry and $H_2S$ starts to deviation the tie-line at 50 GPa (Fig. 1c). This clearly suggests that $H_2S$ is unstable and will decompose to $H_3S$ and sulfur: $3H_2S \rightarrow 2H_3S+S$. Moreover, $H_3S$ remains the most stable stoichiometry up to 300 GPa, the highest pressure examined in this work. In contrast, the calculations show that $H_4S$, $H_5S$, and $H_6S$ stoichiometries are unstable in our study pressure range. Therefore, the dissociation $2H_2S \rightarrow H_4S+S$, $5H_2S \rightarrow 2H_5S+3S$, and $3H_2S \rightarrow H_6S+2S$ are energetically unfavorable.

Form the Fig. 1d it is clearly seen that the $H_2S$ deviates far away the tie-line and the $H_3S$ has the most negative enthalpy of formation at 200 GPa. Therefore, Eremets et al. reported $H_2S$ sample[19] with high $T_c$ of 190 K might decompose to mixture $H_3S$ and S under high pressure. The stable pressure ranges for $H_2S$ and $H_3S$ are predicted in Fig. 2.

The stable structures calculated for each stoichiometry are shown in Fig. 3 and Fig. 4 $H_2S$ is stable in the $P2c$ (20~30 GPa) and $Pc$ (30~50 GPa) structures up to 50

GPa, which are agreement with the recently theoretical reported by Li et al.[15], as shown in Fig. 3a and b. At higher pressures, $H_2S$ decomposes into $H_3S$ and sulfur above 50 GPa, which is consistent with experimental observation that $H_2S$ molecular dissociation near 43 GPa at 150 K[13]. Moreover, in the recently experiment[19], there is a abrupt change in the Raman spectra and $H_2S$ sample starts to conduct at ~50 GPa indicating a structural transition at this pressure.

The high pressure structures, metallization, and superconductivity of $H_3S$ stoichiometry have been elucidated in our previously work.[18] At 20 GPa, $H_3S$ adopts *P*1 structure, which consists of an ordered H-bonded $H_2S$ network and $H_2$ molecular units occupying the interstitial sites, as shown in Fig. 3c. Above 37 GPa, the *Cccm* structure is energetically favored with partial hydrogen bond symmetrization. (Fig.3d). On further compression, $H_2$ units disappear and two intriguing metallic structures with *R*3m and *Im-3m* symmetries (Fig. 3e and 3f) are reconstructive above 111 GPa and 180 GPa, respectively. The *Im-3m* structure is characterized by S atoms located at a simple body centered cubic lattice and H atom located symmetrically between the S atoms and coordination number of the S atom is six.

Although $H_4S$, $H_5S$, and $H_6S$ stoichiometries are only metastable, it is instructive to examine the evolution of the crystal structure comparing with $H_3S$. Selected structures of $H_4S$, $H_5S$, and $H_6S$ at 200 GPa and 300 GPa are shown in Fig. 4. For $H_4S$, it forms orthorhombic structure with *Fmmm* symmetry (16 f.u./cell) at 200 GPa. In this structure, the coordination number of the S atom is six which is the same as *Im-3m* phase of $H_3S$. In addition, there are $H_2$ molecular units between $H_3S$ *Im-3m* blocks. For the case of $H_5S$ at 200 GPa, it forms orthorhombic structure with *Fmmm* symmetry (8 f.u./cell). $H_6S$ adopts orthorhombic structure with *Cmma* symmetry (4 f.u./cell) at 200 GPa. For the structure of $H_4S$, $H_5S$, and $H_6S$ at 200 GPa, there is a common feature that they all consist of $H_3S$ *Im-3m* blocks interbedded $H_2$ molecular units forming sandwich type structure. This suggests the possibility of decomposition to mixture $H_3S$ and $H_2$. In contrast, $H_2$ molecular units disappear in $H_6S$-*Pbcn* structure characteristic as six H-S bonds in the $H_6S$ molecular unit at 300 GPa, as shown in Fig. 4d. It suggests that $H_6S$ may become stable at higher pressures.

## V. Conclusions

In summary, we explore the phase stabilities and the structures of different stoichiometries $H_nS$ (n>1) between 20 and 300 GPa through *ab initio* calculations. The results demonstrate that $H_2S$ decompose to $H_3S$ and sulfur above 50 GPa and $H_3S$ is stable up to 300 GPa. By contrast, the other hydrogen-rich $H_4S$, $H_5S$, and $H_6S$ are unstable in the pressure range we studied. Therefore, $H_2S$ sample with $T_c \approx 190$ K at high pressure might decompose to mixture $H_3S$ and S. Our finding resolve the debate about the pressure-induced decomposition of $H_2S$ for a long time. Further experimental studies of $H_2S$ and pure $H_3S$ at high pressure are still greatly demanded.

## Acknowledgements

This work was supported by the National Basic Research Program of China (No. 2011CB808200), Program for Changjiang Scholars and Innovative Research Team in University (No. IRT1132), National Natural Science Foundation of China (Nos.

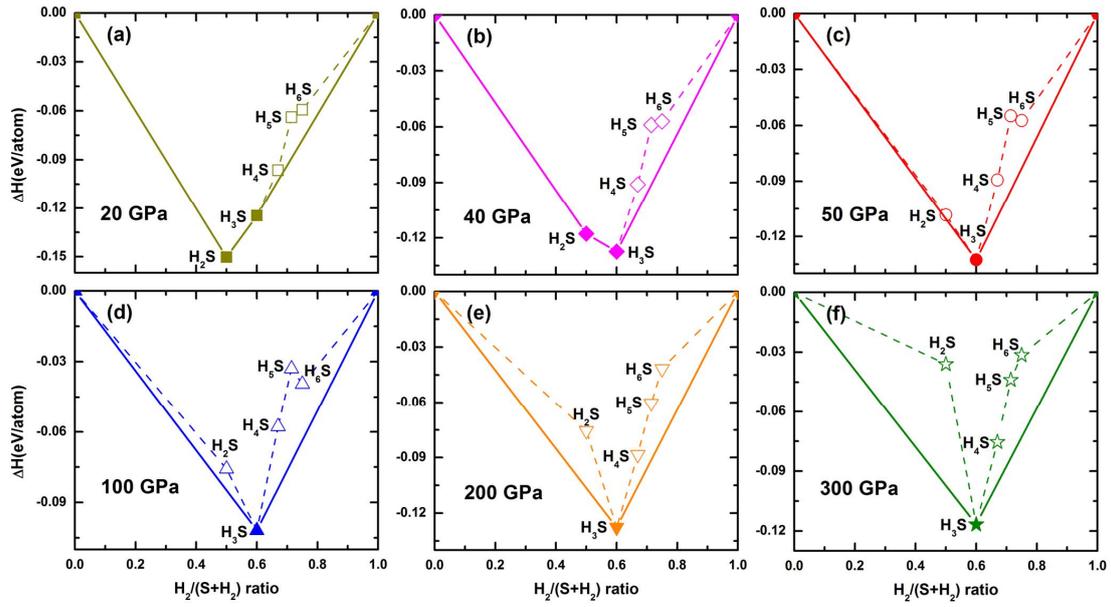

**Fig. 1 | Chemical stabilities of stoichiometric $H_nS$ (n> 1) hydrides under pressure.** Predicted formation enthalpy of $H_nS$ with respect to decomposition into S and $H_2$ under pressure. Dashed lines connect data points, and solid lines denote the convex hull.

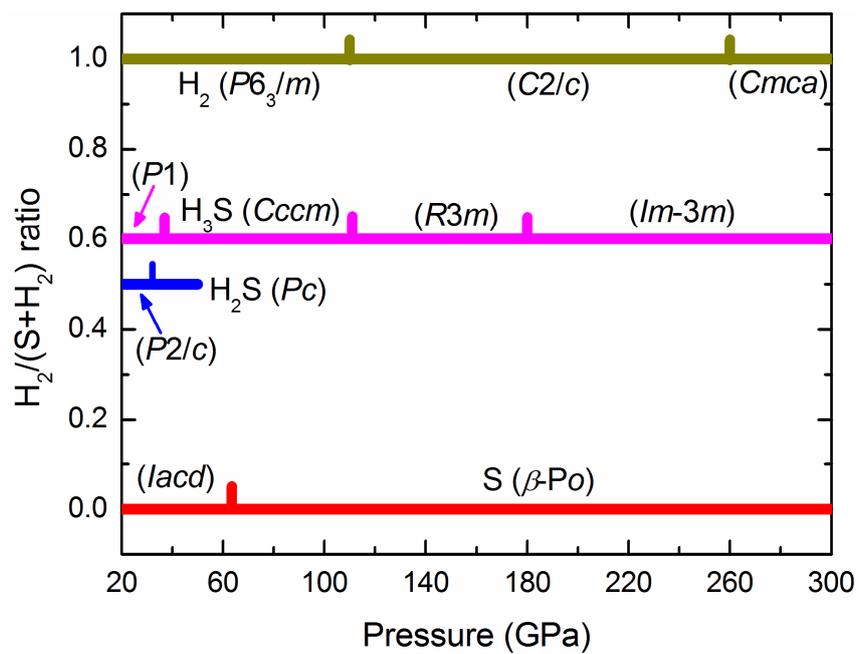

**Fig. 2 | Predicted pressure-composition phase diagram of stoichiometric $H_nS$ (n>1) hydrides.**

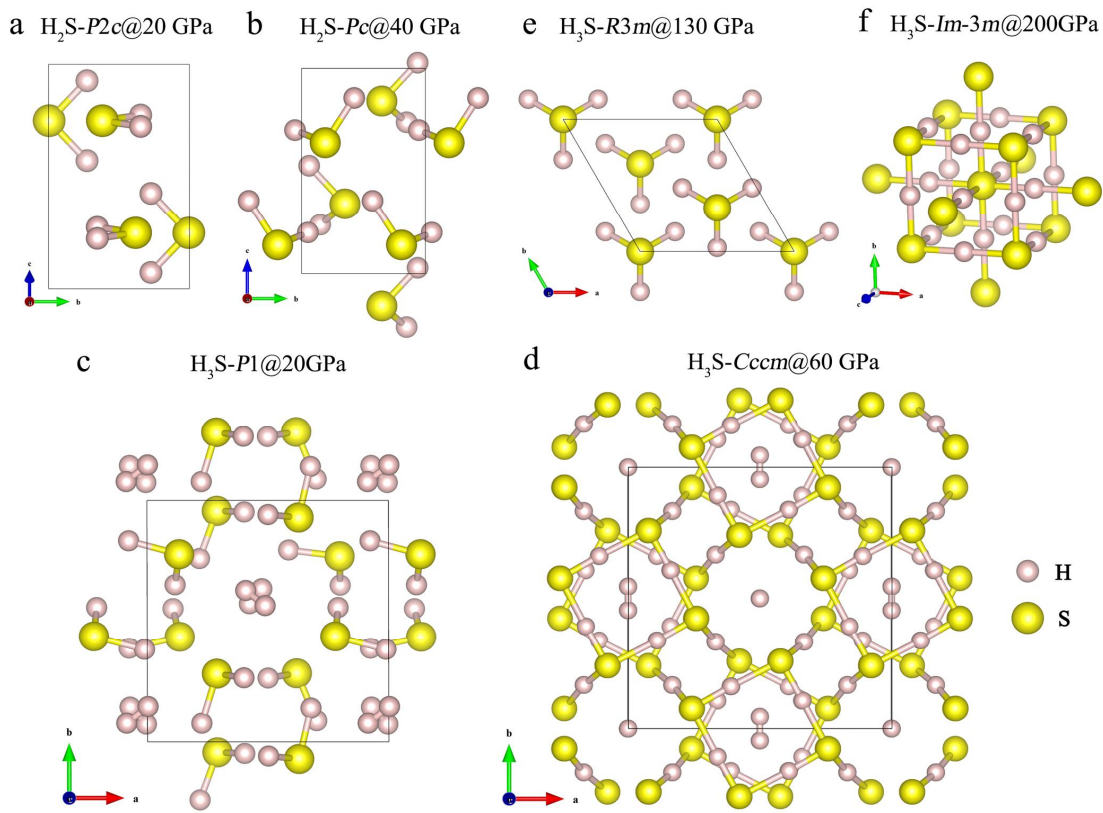

**Fig. 3 | Stable structures of H$_2$S and H$_3$S stoichiometries.** a, H$_2$S at 20 GPa in a *P2c* structure. b, H$_2$S at 40 GPa in a *Pc* structure. c, H$_3$S at 20 GPa in a *P*1 structure. d, H$_3$S at 60 GPa in a *Cccm* structure. e, H$_3$S at 130 GPa in a *R*-3*m* structure. f, H$_3$S at 200 GPa in a *Im*-3*m* structure.

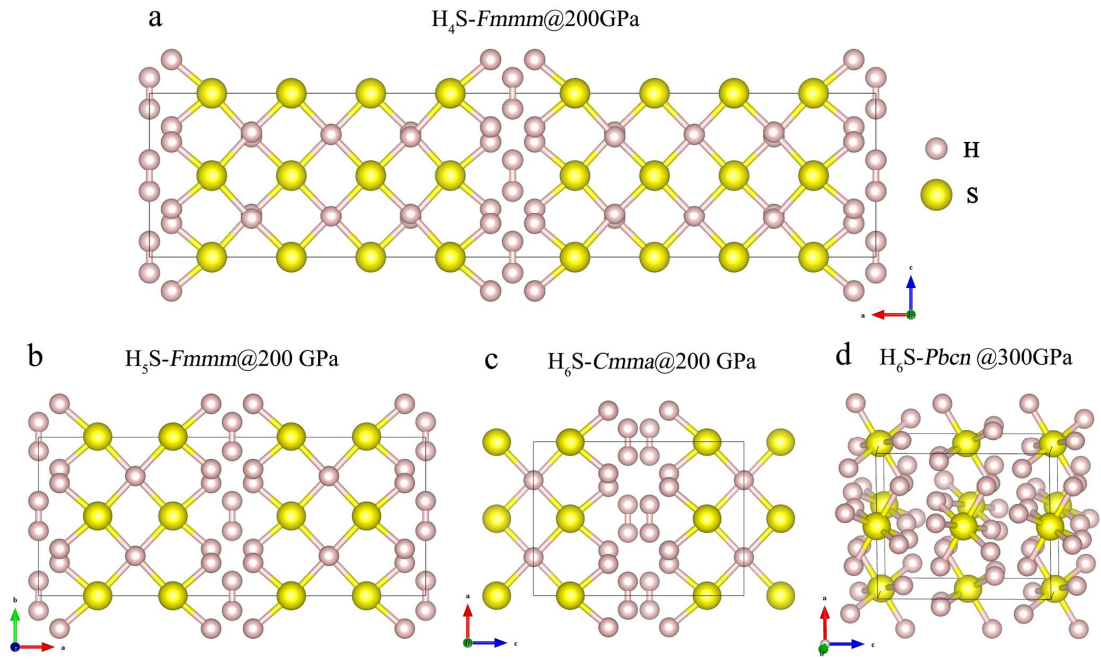

**Fig. 4 | Selected metastable structures of H₄S, H₅S and H₆S stoichiometries.** a, H₄S at 200 GPa in a *Fmmm* structure. b, H₄S at 200 GPa in a *Fmmm* structure. c, H₆S at 200 GPa in a *Cmma* structure. d, H₆S at 300 GPa in a *Pbcn* structure.